# Analysis of the Nutrient Uptake by Roots in Fixed Volume of Soil as Predicted by Fixed Boundary, Moving boundary and Architectural Models

**JUAN C. REGINATO** [1], **JORGE L. BLENGINO** [1] & **DOMINGO A. TARZIA** [2]

[1] *Departamento de Física, Univ. Nacional de Río Cuarto, Ruta Nac.36, Km 603,*

*X5804BYA Río Cuarto, Córdoba, Argentina, jreginato@exa.unrc.edu.ar*

[2] *CONICET-Departamento de Matemática, Univ. Austral, Paraguay 1950, S2000FZF Rosario,*

*Santa Fe, Argentina, DTarzia@austral.edu.ar*

**ABSTRACT**

This work examines the relevance of the one-dimensional models used to study the influx and the cumulative uptake of nutrient by roots. The physical models studied are the fixed boundary model (Barber and Cushman 1981) and an improved version of our moving boundary model (Reginato et al. 2000). A weight averaged expression to compute influx on root surface and a generalized formula to estimate the cumulative nutrient uptake are used. The moving boundary model problem is solved by the adaptive finite element method. For comparison of simulations of influx and cumulative uptake versus observed results six set of data extracted from literature are used. For ions without limitations of availability fixed and moving boundary models produces similar results with small errors. Instead, to low concentrations, the fixed boundary model over predicts while the moving boundary model always produces better results mainly for K. For the P uptake the moving boundary model produces better results only when the concentrations are very low and their predictions are comparable to the obtained by a 3D-architectural model. The obtained improvements would explain any failures of previous models for ions of low availability. Therefore, our model could be a simpler alternative due to its low computational burden.

**INTRODUCTION**

Over the past four decades different mechanistic nutrient uptake models have been developed to simulate nutrient uptake. Usually, these models consist of three basic components (Rengel 1993): (i) solute movement in the soil toward plant roots described by a continuity equation; (ii) nutrient uptake kinetics described by the Michaelis-Menten equation; (iii) nutrient uptake as a result of root growth and inter-root competition by introducing root growth and morphology parameters. Two categories of models have evolved: steady state and transient models (Tinker and Nye 2000). NUTRIENT UPTAKE (Oates and Barber 1987) and NST 3.0 (Claassen et al. 1986) are examples of a transient model with a numerical solution, while SSAND (Li and Comerford 2000) and PCATS (Smethurst et al. 2004) are steady state models. Transient models using numerical solutions are a well-established approach to mechanistic nutrient uptake models (Tinker and Nye 2000). The Barber-Cushman model is a well-known and widely-used model in this category. The model treats the system as two concentric cylinders, where the inner is the root (with constant radius, and no extensions like branching, lateral roots, root hairs or mycorrhizal hypha), whose center is the spatial reference to the soil-root system, with radial orientation. The soil, (assumed homogeneous and isomorphic, with constant moisture content) forms an external cylinder around the root, also with a constant radius. Movement of water and solutes in the soil system is radial to the root only, by mass-flow and diffusion, following Nye and Marriott 1969. Water flow, controlled by the transpiration demand (assumed constant with time), obeys the radial geometry of the system and mass conservation. Nutrient uptake rate is a function of concentration of the ion in question in the soil solution at the root surface, assuming that uptake occurs from a solution only, without interaction with other solutes. Updates to this basic feature include moving boundaries, the external radius (the available soil extent to each root) to account for root growth with time and consequent increase in root density (Reginato et al 2000). NUTRIENT UPTAKE model and NST 3.0 are the personal computer version of the Barber-Cushman model. In 1983 Itoh and Barber developed a submodel to the Barber-Cushman model to include nutrient uptake by root hairs. In 1986 Claassen et al. published NST 1.0 model. In 1987 Oates and Barber published NUTRIENT UPTAKE model. Both were based on the Barber-Cushman model.

Later Claassen and his colleagues developed NST 2.0 and NST 3.0, which were not published in a journal (Claassen, N. Personal communication. 2003, 24 June). NST 3.0 incorporates the Freundlich isotherm into the model so that the buffer power (b) changes as the nutrient concentration in soil solution changes (Steingrobe et al. 2000). Further refinements of the Nye and Marriott derived models consisted to upscale from the root segment to the whole root system, and accounted for root growth (Baldwin et al. 1973; De Willigen et al. 2002). By using upscaling Roose et al. (2001) and Roose & Kirk (2009) provide a fully explicit 'approximate' analytical solution to the Nye–Tinker–Barber model and applied this solution to more complex root branching structure. Roose et al. (2001) showed that the method used to upscale may lead to substantial differences in the predicted uptake of nutrients between their models and NST 3.0 model. When applied to nutrients such as K and P, such models have generally proved quite efficient at predicting the acquisition over time scales of days or weeks for soils receiving high K or P inputs, but almost systematically failed in low input conditions (Brewster et al. 1976; Claassen et al. 1986; Lu and Miller 1994; Mollier et al. 2008; Samal et al. 2010; Schenk and Barber 1980). Under such conditions, those models actually underestimate the observed uptake flux, which suggests that other processes than those accounted for by the models could be operating, and ultimately driving nutrient acquisition. More recent versions started to take into account the effects of fertilizer inputs and nutrient uptake by mycorrhizae (Comerford et al. 2006; Lin and Kelly 2010). However, a comparison of nutrient uptake predictions against experimentally measured values showed that the last version of three process-based models (NST 3.0, SSAND, and PCATS) largely underestimated P uptake for three woody plant species, except under large P fertilizer additions for the transient state model NST 3.0 developed by Claassen and co-workers (e.g.Claassen et al. 1986; Steingrobe et al. 2000). This pattern showed that including mycorrhizal uptake in the simulations was not sufficient to predict accurately nutrient uptake under the low nutrient concentrations. These results suggested that rhizospheric effects, not yet taken into account in these models, could be carried out to improve their predictive ability. Further 3D root system architecture models were RootTyp, SimRoot, ROOTMAP, SPACSYS, R-SWMS, and RootBox (Dunbabin et al., 2013) and they are being used to study how specific root traits affect the uptake of a variety of soil resources such as nitrogen, phosphorus, and water. Limitations of current models also infer the future directions in research. Though root system architecture modeling is being used to understand plant uptake of soil resources, there are problems with upscaling from single roots to the whole system. Current models do not include rhizosphere processes or soil microorganisms that are known to be important for resource uptake and root growth. From other point of view, heat and mass transfer with phase change problems such as evaporation, condensation, freezing, melting, sublimation, have wide application in separation processes, food technology, heat and mixture migration in soils and grounds, etc. They have been studied in the last century due to their wide scientific and technological applications [Alexiades-Solomon, 1993; Cannon, 1984; Carslaw-Jaeger, 1959; Crank, 1984; Lunardini, 1991]. This kind of problem are known in the literature as free or moving boundary problems depending if the interface is unknown or known a priori. A large bibliography and a review of explicit solutions on free and moving boundary problems for the heat-diffusion equation were given in [Tarzia, 2000, 2011]. Due to the non-linearity of the problems, solutions usually involve mathematical difficulties and analytical methods can only be developed for idealized systems or for those with plain structure and constant properties. Numerical methods, instead, allow considering the structural and compositional characteristics through detailed models. The methodology of free and moving boundary problems have been also applied to agronomic problems. Thus, there are previous papers in this area for which the nutrient uptake has been implicitly modeled by moving boundary problems, for example, Abbes et al., 1996, Huguenin and Kirk, 2003. Explicit one-dimensional free and moving boundary model applied to root growth and nutrient uptake was presented in Reginato et al., 1990, 2000, Jonard et al., 2010. The goal of this paper is to consider an improved version of the moving boundary model applied to uptake of ions of low, medium and high availability by roots. In particular, we will revise and compare the uptake of ions through model NST 3.0, and the moving boundary model. For both models, we use a new generalized cumulative uptake formula and the moving boundary model is solved by the adaptive finite elements method. Moreover, we also compare the cumulative uptake predicted by a more complex root branching system model with ones obtained by our moving boundary model.

*The moving boundary model*

This model is based on the same assumptions formulated by the Barber-Cushman model but, now, the model incorporates a new boundary condition for root competition (among roots of root system) which represents the net flux on the moving boundary R(t). This moving boundary is given by the instantaneous half distance between roots axis which is the result of the root length variation. Thus, moving boundary R(t) is a function of the instantaneous root length $\ell$ (t) which is a known function of time. A representation of the new condition can be visualised assuming a fixed volume of soil in which the root system is distributed like a homogenous piling up by roots, i.e., we propose an idealized total root system submerged in a fixed volume of soil (pots) instead of a single root in an infinite volume of soil (Figure 1). The conditions of humidity, light and temperature are assumed to be controlled (as in a growth chamber). Based on these assumptions and using root length density as a function of t, R(t) (the moving boundary), the following set of equations and boundary conditions in cylindrical coordinates are used:

$$\left(\phi+b\right)\frac{\partial C}{\partial t}=\frac{\phi D}{r}\frac{\partial}{\partial r}\left(r\frac{\partial C}{\partial r}\right)+\frac{vs_o}{r}\frac{\partial C}{\partial r}, \qquad s_o<r<R(t), \quad t>0 \tag{1}$$

$$C(r,0)=C_o(r), \qquad s_o<r<R_o \tag{2}$$

$$D\phi\frac{\partial C(s_o,t)}{\partial r}+vC(s_o,t)=\frac{J_m\left[C(s_o,t)-C_u\right]_+}{K_m+\left[C(s_o,t)-C_u\right]_+}, \quad t>0 \tag{3}$$

$$D\phi\frac{\partial C(R(t),t)}{\partial r}+\frac{vs_o}{R(t)}C(R(t),t)=0, \qquad t>0 \tag{4}$$

where the moving boundary is given by:

$$R(t)=\sqrt{\frac{\ell_o}{\ell(t)}(R_o^2-s_o^2)+s_o^2}, \quad t>0 \tag{5}$$

where *r* is the radial distance from the axis of the root [cm]; *t* is the time [s]; b is the buffer power [dimensionless]; D is the diffusion coefficient in soil [cm$^2$ s$^{-1}$] ($=D_f f$ , where $D_f$ is the diffusion coefficient in free liquid and f is a tortuosity factor); $s_o$ is the root radius [cm]; v is the effective velocity of flux solution [cm s$^{-1}$]; $R_o$ is the initial half distance among root axis [cm]; $J_m$ is the maximum influx [mol cm$^{-2}$ s$^{-1}$]; $K_m$ is the concentration for which the influx is $J_m$ /2 [mol cm$^{-3}$]; $C_u$ is the threshold concentration below which influx stops [mol cm$^{-3}$]; R(t) is the half distance among roots axis [cm]; $C_o(r)$ is the initial concentration profile in [$s_o$, $R_o$] [mol cm$^{-3}$] and $\ell(t)$ is the known root length as a function of time [cm] (the known law of root growth which can be linear, exponential or sigmoid; in the computed results we have used for some case the linear growth $\ell(t) = \ell_o + kt$ with units of k in [cm s$^{-1}$] and the exponential growth defined by $\ell(t) = \ell_o e^{kt}$ with units of k in [s$^{-1}$]. We denote $x_+$ as the part positive of x defined by $x_+ = \text{Max}(0;x)$. Equation (1) is the equation of diffusive and convective transport of ions in soil and condition (2) corresponds to the initial profile of concentrations. Condition (4) represents a null flux on the moving limit of not-transference or instantaneous half mean distance between roots R(t). We remark that the null flux condition imposed in this paper by equation (4) is a more realistic condition and a corrected version with respect to the similar one used in our previous model (Reginato et al., 2000). Condition (3) represents the mass balance on the root surface and the expression (5) represents the moving boundary R(t) as a function of the instantaneous root length $\ell(t)$. Expression (5) for the moving boundary is an improved version of a similar condition used in our previous model and introduces minor error in the computational algorithm designed to solve the problem. Unlike the expression proposed for R(t) in the previous version of our model which was based on considerations of constant volume of soil including roots (If the total volume of root plus soil remains constant then the amount of soil available to root will not be constant over time, i.e., the root grows at the expense of the decrease of the volume of soil), is now considered a constant volume of soil (See Appendix I. A schematic diagram of the problem (1) – (5) is shown in the Figure 2.

The solution of problem (1) – (5) is obtained by the application of the finite element method by using a dimensionless formulation through the following change of variable (similar to the one proposed by Roose (2009), but now scaling the difference of coordinates (r-so) by the difference of coordinates (R(t)-so), i.e., transforming the variable interval (so,R(t)) in a fixed interval (0,1) for all t > 0:

$$C^*(r^*,t^*) = \frac{C(r,t)}{K_m}, \qquad r^* = \frac{r - s_0}{R(t) - s_0}, \qquad t^* = \frac{D\phi}{(\phi + b)R_o^2}t \quad . \tag{6}$$

Thus, we obtain the following dimensionless problem in a fixed domain (See Appendix II):

$$C_{t^*}^* = \left(\frac{R_o}{s_o}\right)^2 \frac{C_{r^*r^*}^*}{\left(R^*(t^*)-1\right)^2} + \frac{C_{r^*}^*}{R^*(t^*)-1}\left[(1+Pe)\left(\frac{R_o}{s_o}\right)^2 \frac{1}{1+r^*\left(R^*(t^*)-1\right)} + r^*\frac{dR^*}{dt^*}\right],$$

$$0 < r^* < 1, \ \ t^* > 0 \tag{7}$$

$$C^*(r^*,0) = C_o^*(r^*), \qquad 0 < r^* < 1 \tag{8}$$

$$\frac{C_{r^*}^*(0,t^*)}{R^*(t^*)-1} = \lambda \frac{\left[C^*(0,t^*)-C_u^*\right]_+}{1+\left[C^*(0,t^*)-C_u^*\right]_+} - P_e C^*(0,t^*), \quad t^* > 0 \tag{9}$$

$$\frac{C_{r^*}^*(1,t^*)}{R^*(t^*)-1} + P_e \frac{C^*(1,t^*)}{R^*(t^*)} = 0, \qquad t^* > 0 \tag{10}$$

where:

$$R^*(t^*) = \frac{1}{s_o} R\left(\frac{(\phi+b)R_o^2}{D\phi}t^*\right) = \begin{cases} \sqrt{\dfrac{1+\left[(R_o/s_o)^2-1\right]}{1+\dfrac{k(\phi+b)R_o^2}{D\phi\ell_o}t^*}} & \text{if } \ell(t) = \ell_o + kt \\ \\ \sqrt{\dfrac{1+\left[(R_o/s_o)^2-1\right]}{e^{\frac{k(\phi+b)R_o^2}{D\phi}t^*}}} & \text{if } \ell(t) = \ell_o e^{kt} \end{cases}, t^* > 0 \tag{11}$$

and Pe (Peclet number), $\lambda$ and $C_u^*$ are dimensionless parameters defined by:

$$Pe = s_o v / D\phi, \qquad \lambda = \frac{J_m s_o}{D\phi K_m}, \qquad C_u^* = \frac{C_u}{K_m} \tag{12}$$

by considering:

$$C_{t^*}^* = \frac{\partial C^*}{\partial t^*}, \quad C_{r^*}^* = \frac{\partial C^*}{\partial r^*}, \quad C_{r^*r^*}^* = \frac{\partial^2 C^*}{\partial r^{*2}} \tag{13}$$

We remark that the moving boundary problem (1)-(5) was transformed in a fixed boundary problem, but the partial differential equation (7) now takes into account the moving boundary R(t) (or $R^*(t^*)$ in the dimensionless domain).

We remark that, the dimensionless equation (7) reduces to the dimensionless equation of Roose (Roose et al., 2009) when $R(t) = R_o$ = constant (fixed domain) and the root radius $s_o$ is small with respect to the half distance among roots $R_o$. Thus, our moving boundary formulation is a generalization of the fixed boundary model (Cushman, 1979, Roose, 2001) (See Appendix III). The solution of the nutrient uptake problem (7) – (11) in an inmobilized domain is obtained by a more stable and efficient numerical method that satisfies the mass balance among the ions taken by the root system and the ions remaining in soil. Thus, we apply the finite element method by using the FlexPDE software (http://www.pdesolutions.com, Schnepf et al., 2002). Once the influx values on the root surface are obtained we estimate the cumulative nutrient uptake by our growing root system by the following generalized formula valid for any range of concentrations (See Appendix IV):

$$\Delta U = U(t_f) - U(t_i) = 2\pi s_o \int_{t_i}^{t_f} J(t)\ell(t)d\tau \qquad (14)$$

where J is given in mol cm$^{-2}$ s$^{-1}$, $\ell(t)$ in cm and U in moles. The influx J(t) is given by

$$J(t) = J_m \left[ C(s_o, t) - C_u \right] / \left[ K_m + \left( C(s_o, t) - C_u \right) \right].$$

Moreover, and based in this generalized formula; we define a weight averaged influx which is consistent with the experimental William´s formula (Williams, 1946, 1948). This averaged influx is given by (See Appendix V):

$$\overline{J} = \frac{\int_{t_i}^{t_f} J(t)\ell(t)dt}{\int_{t_i}^{t_f} \ell(t)dt} \qquad (15)$$

where $\overline{J}$ is given in mol/cm$^2$-s. This weight averaged influx is more realistic because takes into account the temporal contribution of root length to the influx. In the case of constant influx J(t)=J, the averaged influx given by (15) coincides with the temporal averaged influx

$$\frac{1}{(t_f - t_i)} \int_{t_i}^{t_f} J(t)dt .$$

**The Simulations**

For comparison of simulations of influx on root surface and cumulative uptake versus observed data we use six set of input data sets extracted from literature. First, we compute the influx on root surface of Cd by maize, sunflower, flax and spinach for two levels of concentration (Stritsis et al. 2014) and the results are shown in Table 1 and Figure 3. The influxes obtained by our MB-FE model are averaged by using the formula (15). Second, we compute the cumulative K and P uptake by pine seedling (Kelly et al. 1992) and the cumulative $NO_3^-$ uptake by wheat; both to high nutrient concentrations (Jia-Xiang et al. 1991) and the results are shown in Tables 2 and 3. Third, we compute the influx on root surface and cumulative K uptake by maize, wheat and sugar beet for low K soil and soil with K addition (Samal et al. 2010) and the results are shown in Tables 4, 5 and 6. Fourth, we compute the influx on root surface and cumulative P uptake by peanut for low, intermediate and high soil concentrations (Singh et al. 2003) and the results are shown in Tables 7, 8 and 9. Finally, in order to verify the reliability of our moving boundary model and a 3D-dimensional architectural model we compute the P uptake by wheat to low concentrations (Heppel et al. 2014) and the results are shown in Figures 4, 5 and 6. From now on, we denote the simulations as:

FB-NST 3.0: Original fixed boundary model NST 3.0,

FB-NST 3.0*: Fixed boundary model NST 3.0 with generalized nutrient uptake formula (14),

MB-FE: Moving boundary model solved by finite element method with generalized nutrient uptake formula (14).

**RESULTS AND DISCUSSION**

From Table 1, we conclude that MB-FE model predicts the average influx on root surface better than the FB-NST 3.0 always for different plants for two levels of concentrations. From Figure 3, we obtain that the correlation factor for the linear regression among the observed and predicted influxes is 0.73 for FB-NST 3.0 model and 0.89 for the MB-FE model. From Table 2 for high concentrations and immobile ions (K, P), very low Peclet numbers and low variation of root length density (arboreal species) FB-NST 3.0, FB-NST 3.0* and MB-FE methods produce similar results with small errors for the influx on root surface and the cumulative nutrient uptake. From Table 3 for high concentrations and mobile ions ( $NO_3^-$ ), very low Peclet numbers and moderate variation of root length density FB-NST 3.0, FB-NST 3.0* and MB-FE models produces similar results for the influx on root surface and the cumulative nutrient uptake. From Table 4 for low and increasing level of P concentration the fixed (original NST 3.0 and NST3.0*) and moving MB-FE models over predicting always except in the soil without P addition (low concentrations). In this last case, MB-FE under predict with an acceptable error. From Table 5, we conclude that to low concentrations both models FB-NST 3.0 (with temporal average influx) and MB-FE (with weight average influx) under predict the average influx. For increasing level of P addition both models over predict the average influx. In Table 6, we show the average predicted/observed ratio for influx and cumulative uptake obtained by NST 3.0, NST 3.0* and MB-FE models. We conclude that MB-FE predict better the average predicted/observed ratio for the influx and the cumulative nutrient uptake. From Table 7, we conclude that for ions as K the MB-FE model is the best numerical method to compute the cumulative K uptake on a low K soil with K addition and without K addition for all cases. From Table 8, we conclude that in almost all cases, FB-NST 3.0 produces better predicted influxes except for maize without K addition. From Table 9, we conclude that the average/predicted ratio of influx on root surface are similar although for the average predicted/observed ratio of cumulative K uptake our MB-FE model presents a remarkable improvement in the predictions showing an average predicted/observed ratio of cumulative K uptake (1.25) in front of those obtained by NST 3.0 (15.6) and NST 3.0* (10.9). From Figure 4, we conclude that our MB-FE model predict better the final cumulative P uptake at 10 days. Although cumulative uptake predicted by the 3D model best fit to the experimental data we remark that this setting is obtained by choosing the best branching mode that fits to the experimental curve. Thus, the 3D model depends of the branching model while our MB-FE model is not depending but both models produce the same final value for the cumulative uptake (see Figure 5)

The reason for which the MB-FE model is better than the other schemes (FB-NST 3.0 and FB-NST $3.0^*$) is that these last methods do not satisfy the mass balance among the ions taken by root and the ions remaining in soil. For the finite element method the ions remaining in soil were calculated by the following expression:

$$N(t) = 2\pi(1+b)\ell(t)\int_{s_o}^{R(t)} rC(r,t)dr$$

which, in the dimensionless interval (0, 1), is transformed in:

$$N(t) = 2\pi(1+b)s_o^3 K_M \ell^*(t^*)\int_0^1 \left[1 + r^*\left(R^*(t^*)-1\right)\right]\left[R^*(t^*)-1\right]C^*(r^*,t^*)dr^* \ .$$

The mass balance for the program NTS 3.0 is not considered here because, obviously, to compute the ions remaining in soil, the operation must be done with concentration profiles as a function of time which has been calculated in fixed domain, but this result must be compared with the cumulative uptake by a growing root, i.e., which has been calculated by integration in a variable domain (the root growing in an volume of soil available that is variable in the time). We remark that NTS 3.0 model satisfies the mass balance only in the case when the roots are not growing. That is, the ion transport dynamics in a fixed soil domain is only consistent with a fixed root domain; thus, our MB-FE model is a more realistic description because the nutrient uptake by a variable root domain is consistent with the ions transport dynamics in a variable soil domain. Figure 6 show the mass balance for the results obtained by the finite elements method. The calculus was done with data extracted from Heppel (2014) for the P uptake by wheat to low concentration. Loss mass is the difference among ions remaining in soil and cumulative uptake, and it is produced by the implicit errors propagation in the numerical method employed. Figure 7 illustrate the differences between the fixed boundary model and our moving boundary model. Graph **a** shows the ion concentration on root surface $C(s_o,t)$, Graph **b** shows the influx on root surface $J(s_o,t)$, Graph **c** shows the instantaneous nutrient uptake $2\pi s_o J(t)\ell(t)$, and Graph **d** shows the cumulative nutrient uptake obtained being all the graphics for the two models. Graph **e** shows the mass balance for the MB-FE model. The data used were taken from Samal (2010) for the K uptake by maize without addition of K. For comparison of Figure 6 and Figure 7e the loss mass is more pronounced for the K uptake by maize although this is a known issue of finite element method when high gradients of concentration are present.

The obtained improvements by our model are mainly due to three factors: a) the use of a generalized formula for the cumulative nutrient uptake, b) the influxes obtained by the moving boundary model and the cumulative uptake, which are obtained through integration in a variable domain, while for the fixed boundary model the influxes are obtained in a fixed domain and the cumulative uptake by integration is on a variable domain, c) the use of a numerical method (finite element method) that ensures the balance of mass among the absorbed ions and the remaining ions in soil while the finite difference method does not satisfy it. Finally, in the light of these findings, conclusions drawn by previous papers (Hinsinger, 2011) could be reinterpreted and our model could be included in larger field/catchment/climate scale models something which is not practically possible with the 3D numerical simulations due to their high computational burden. Using the model presented in this paper we have been able to give a new insight into the discrepancies and inaccuracies present in previous models. However, to clarify and verify this, more accurate simultaneous measurements of nutrient and water uptake together with root branching structure development should be carried out. With this article we have developed an alternative and consistent form to compute the absorption of nutrients by the root system, thus allowing the results of the experiments can be analyzed in an easily interpretable way and resolve some problems with the up scaling of single roots to the whole system through the mass balance. Moreover, our moving boundary model open a new perspective for the study of nutrient uptake in variable soil volume (to field) in order to solve the differences between predicted uptake in fixed and variable soil volume (Silberbush, 2013), which it is not possible with fixed boundary models and 3D architectural models based in concentrations and influxes calculated in fixed domains.

**APPENDIX I.** Expression (5) is obtained assuming a soil volume constant, i.e.:

$$V_{soil}(0) = V_{soil}(t), \quad \forall t > 0 \quad \Leftrightarrow \quad \pi\left(R_o^2 - s_o^2\right)\ell_o = \pi\left(R^2(t) - s_o^2\right)\ell(t)$$

$$\Leftrightarrow \quad R^2(t) = \left(R_o^2 - s_o^2\right)\frac{\ell_o}{\ell(t)} + s_o^2 \quad \Leftrightarrow \quad R(t) = \sqrt{\left(R_o^2 - s_o^2\right)\frac{\ell_o}{\ell(t)} + s_o^2} \quad .$$

**APPENDIX II:** Taking into account the change of variables (6) we obtain:

i)
$$r = s_o\left[1 + r^*\left(R^*(t^*)\right) - 1\right]$$

ii)
$$\begin{cases} R(t) = s_o R^*(t^*) \\ \ell(t) = s_o \ell^*(t^*) \end{cases}$$

iii)
$$\dot{R}(t) = \frac{dR(t)}{dt} = \frac{s_o D\phi}{(\phi + b)R_o^2}\frac{dR^*(t^*)}{dt^*}$$

iv)
$$C_t(r,t) = K_M\left[\frac{D\phi}{(\phi + b)R_o^2}C_{t^*}^*(r^*,t^*) - \frac{r^*\dot{R}(t)}{R(t) - s_o}C_{r^*}^*(r^*,t^*)\right]$$

v)
$$C_r(r,t) = \frac{K_M}{R(t) - s_o}C_{r^*}^*(r^*,t^*)$$

vi)
$$C_{rr}(r,t) = \frac{K_M}{\left(R(t) - s_o\right)^2}C_{r^*r^*}^*(r^*,t^*)$$

Then, the partial differential equation (1) is transformed in (7). The boundary condition (2) is transformed in (8) where

$$C_o^*(r^*) = \frac{C_o\left[s_o + r^*\left(R_o - s_o\right)\right]}{K_M} \quad .$$

The boundary condition (3) at $r = s_o$ is transformed in (9) where the dimensionless parameter $\lambda$ is given by (12) and $C_u^* = C_u / K_M$

The boundary condition (4) at the moving boundary $r = R(t)$ is transformed in (10) where the dimensionless parameter Pe is given by (12).

The expression (5) for the instantaneous half distance between roots axis is transformed in (11) according to the choice of the root length growth law (linear o exponential type). We deduce for the growth linear case $\ell(t) = \ell_o + kt$ the following expressions:

$$\ell^*(t^*) = \frac{\ell_o + kt}{s_o} = \frac{\ell_o}{s_o}\left(1 + \frac{k(\phi+b)R_o^2}{D\phi\ell_o}t^*\right)$$

$$R^*(t^*) = \frac{R(t)}{s_o} = \sqrt{\frac{\ell_o}{\ell(t)}\left(\frac{R_o^2}{s_o^2}-1\right)+1} = \sqrt{\frac{\ell_o}{s_o\ell^*(t^*)}\left[\left(\frac{R_o}{s_o}\right)^2-1\right]+1}$$

$$= \sqrt{\frac{\ell_o}{s_o\dfrac{\ell_o}{s_o}\left(1+\dfrac{k(\phi+b)R_o^2}{D\phi\ell_o}t^*\right)}\left[\left(\frac{R_o}{s_o}\right)^2-1\right]+1}$$

$$= \sqrt{1+\frac{\left[\left(\dfrac{R_o}{s_o}\right)^2-1\right]}{\left(1+\dfrac{k(\phi+b)R_o^2}{D\phi\ell_o}t^*\right)}}$$

And for the exponential growth case $\ell(t) = \ell_o e^{kt}$ we obtain the following expressions:

$$\ell^*(t^*) = \frac{\ell_o e^{kt}}{s_o} = \frac{\ell_o}{s_o}e^{\frac{k(\phi+b)R_o^2}{D\phi\ell_o}t^*}$$

$$R^*(t^*) = \frac{R(t)}{s_o} = \sqrt{\frac{\ell_o}{\ell(t)}\left(\frac{R_o^2}{s_o^2}-1\right)+1} = \sqrt{\frac{\ell_o}{s_o\ell^*(t^*)}\left[\left(\frac{R_o}{s_o}\right)^2-1\right]+1}$$

$$= \sqrt{\frac{\ell_o}{s_o\dfrac{\ell_o}{s_o}e^{\frac{k(\phi+b)R_o^2}{D\phi\ell_o}t^*}}\left[\left(\frac{R_o}{s_o}\right)^2-1\right]+1}$$

$$= \sqrt{1+\frac{\left[\left(\dfrac{R_o}{s_o}\right)^2-1\right]}{e^{\frac{k(\phi+b)R_o^2}{D\phi\ell_o}t^*}}}$$

**APPENDIX III:** If we take into account the partial differential equation (7) in a fixed domain, that is $R(t) = R_o = $ constant ($R^*(t^*) = R_o/s_o$), then equation (7) reduces to

$$C_{t^*}^* = \frac{C_{r^*r^*}^*}{\left(1-\dfrac{s_o}{R_o}\right)^2} + \frac{(1+Pe)C_{r^*}^*}{\left(1-\dfrac{s_o}{R_o}\right)\left[\dfrac{s_o}{R_o}+r^*\left(1-\dfrac{s_o}{R_o}\right)\right]} \quad . \tag{16}$$

Moreover, if the root radius $s_o$ is small with respect to the half distance between roots $R_o$ ($s_o/R_o \ll 1$) then (16) reduces to

$$C_{t^*}^* - Pe \frac{C_{r^*}^*}{r^*} = C_{r^*r^*}^* + \frac{1}{r^*}C_{r^*}^* = \frac{1}{r^*}\frac{\partial}{\partial r^*}\left(r^*\frac{\partial C^*}{\partial r^*}\right) \tag{17}$$

that is equals to equation (5) of Roose and Kirk (2009, page 259).

**APPENDIX IV:** Simulations of transport and nutrient uptake by roots generate a set of influxes data at the interface root-soil. From these influxes the cumulative nutrient uptake is calculated integrating on a variable and growing domain (the root). The formulas used for these integrals up to now are the following: The Claasen-Barber formula (1976)

$$U_{CB} = 2\pi s_o \ell(t_i)\int_{t_i}^{t_f} J(t)dt + 2\pi s_o \int_{t_i}^{t_f}\left[\int_{t_i}^{\tau} J(t)dt + \right]\dot{\ell}(\tau)d\tau , \tag{18}$$

the Cushman formula (1979)

$$U_C = 2\pi s_o \ell(t_i)\int_{t_i}^{t_f} J(t)dt + 2\pi s_o \int_{t_i}^{t_f}\left[\int_{t_i}^{t_f-\tau} J(t)dt + \right]\dot{\ell}(\tau)d\tau , \tag{19}$$

and the Reginato-Tarzia formula (2002)

$$U_{RT} = 2\pi s_o \ell(t_i)\int_{t_i}^{t_f} J(t)dt + 2\pi s_o \int_{t_i}^{t_f}\left[\int_{\tau}^{t_f} J(t)dt + \right]\dot{\ell}(\tau)d\tau . \tag{20}$$

The formula (20) for $U_{RT}$ can be, integrating by parts**,** simplified in order to obtain:

$$
\begin{aligned}
U_{RT} &= 2\pi s_o \ell(t_i)\int_{t_i}^{t_f} J(t)dt + 2\pi s_o \int_{t_i}^{t_f}\underbrace{\left[\int_{\tau}^{t_f} J(s)ds\right]}_{V}\underbrace{\dot{\ell}(\tau)}_{\dot{U}}d\tau\\
&= 2\pi s_o \ell(t_i)\int_{t_i}^{t_f} J(t)dt + 2\pi s_o \left[\ell(\tau)\left(\int_{\tau}^{t_f} J(s)ds\right)\Big|_{\tau=t_i}^{\tau=t_f} + \int_{t_i}^{t_f} J(t)\ell(t)dt\right]\\
&= 2\pi s_o \ell(t_i)\int_{t_i}^{t_f} J(t)dt + 2\pi s_o \left[0.\,\ell(t_f) - \ell(t_i)\left(\int_{t_i}^{t_f} J(t)dt\right) + \int_{t_i}^{t_f} J(t)\ell(t)d\tau\right]\\
&= 2\pi s_o \ell(t_i)\int_{t_i}^{t_f} J(t)dt - 2\pi s_o \ell(t_i)\int_{t_i}^{t_f} J(t)dt + 2\pi s_o \int_{t_i}^{t_f} J(t)\ell(t)dt\\
&= 2\pi s_o \int_{t_i}^{t_f} J(t)\ell(t)d\tau
\end{aligned}
\tag{21}
$$

where $t_i$ and $t_f$ are the initial and final times respectively, $s_o$ is the root radius, $\ell(t_i)$ is the initial root length, $J(t)$ is the influx on the root surface as a function of the time t and $\dot{\ell}(t)$ is the rate of root growth as a function of time. The product $2\pi s_o J(t)\ell(t)$ is the instantaneous nutrient uptake by unit of time.

The last formula (21) can be also obtained from basic physics principles owing to $J(t)$ is the amount of nutrient by unit of time and unit of area that the root takes. That is to say that if we multiply J(t) by the instantaneous lateral root area ( $2\pi s_o\,\ell(t)$ ) we get the instantaneous amount of nutrients per unit of time that takes the root, integrating this versus time we get the proposed formula (21).

In order to determine the range of validity of the formulas (18), (19) or (20) (or his equivalent simplified expression given by (21)) a mass balance is performed according to the following procedure: if we subdivide the integration interval $(t_i, t_f)$ from $t_i$ until an intermediate time $t$ and from this time until the final time $t_f$ then the formulas satisfy the mass balance when the cumulative nutrient uptake $U$ verify the following equality:

$$U(t_i, t_f) = U(t_i, t) + U(t, t_f) = \text{SUM CUMULATIVE UPTAKE,} \quad \text{for all } t_i < t < t_f \ . \quad (22)$$

First, we analyse the Claassen-Barber formula (18). In order to make easier this study, we first simplify the referred formula integrating by parts:

$$
\begin{aligned}
U_{CB} &= 2\pi s_o\left[ \ell(t_i)\int_{t_i}^{t_f} J(s)ds + \int_{t_i}^{t_f} \underbrace{\left[\int_{t_i}^{\tau} J(s)ds +\right]}_{f(\tau)} \overset{\bullet}{\ell}(\tau)d\tau \right] \\
&= 2\pi s_o\left[ \ell(t_i)\int_{t_i}^{t_f} J(s)ds + f(\tau)\ell(\tau)\Big|_{\tau=t_i}^{\tau=t_f} - \int_{t_i}^{t_f} \underset{J(\tau)}{f'(\tau)}\,\ell(\tau)d\tau \right] \\
&= 2\pi s_o\left[ \ell(t_i)\int_{t_i}^{t_f} J(s)ds + f(t_f)\ell(t_f) - \underset{=0}{f(t_i)\,\ell(t_i)} - \int_{t_i}^{t_f} J(\tau)\ell(\tau)d\tau \right] . \quad (23) \\
&= 2\pi s_o\left[ \ell(t_i)\int_{t_i}^{t_f} J(s)ds + \ell(t_f)\int_{t_i}^{t_f} J(s)ds - \int_{t_i}^{t_f} J(\tau)\ell(\tau)d\tau \right] \\
&= 2\pi s_o\int_{t_i}^{t_f}\left[ \ell(t_i) + \ell(t_f) - \ell(\tau) \right] J(\tau)d\tau
\end{aligned}
$$

Now, in order to decide if $U_{CB}$ verifies (22) we compute:

$$\frac{U_{CBS}(t_i, T) + U_{CBS}(T, t_f) - U_{CBS}(t_i, t_f)}{2\pi s_o} =$$

$$= \int_{t_i}^{t} \left[\ell(t_i) + \ell(t) - \ell(\tau)\right] J(\tau) d\tau + \int_{t}^{t_f} \left[\ell(t) + \ell(t_f) - \ell(\tau)\right] J(\tau) d\tau$$

$$- \int_{t_i}^{t_f} \left[\ell(t_i) + \ell(t_f) - \ell(\tau)\right] J(\tau) d\tau$$

$$= \left[\ell(t_i) + \ell(t)\right] \int_{t_i}^{t} J(\tau) d\tau - \int_{t_i}^{t} J(\tau)\ell(\tau) d\tau + \left[\ell(t) + \ell(t_f)\right] \int_{t}^{t_f} J(\tau) d\tau$$

$$- \int_{t}^{t_f} J(\tau)\ell(\tau) d\tau - \left[\ell(t_i) + \ell(t_f)\right] \int_{t_i}^{t_f} J(\tau) d\tau + \int_{t_i}^{t_f} J(\tau)\ell(\tau) d\tau \qquad , \qquad (24)$$

$$= \left[\ell(t) - \ell(t_f)\right] \int_{t_i}^{t} J(\tau) d\tau + \left[\ell(t) - \ell(t_i)\right] \int_{t}^{t_f} J(\tau) d\tau$$

$$= \ell(t) \int_{t_i}^{t} J(\tau) d\tau - \ell(t_f) \int_{t_i}^{t} J(\tau) d\tau - \ell(t_i) \int_{t}^{t_f} J(\tau) d\tau$$

$$= \left[\ell(t) - \ell(t_i)\right] \int_{t}^{t_f} J(\tau) d\tau - \left[\ell(t_f) - \ell(t)\right] \int_{t_i}^{t} J(\tau) d\tau$$

that is

$$U_{CB}(t_i, t) + U_{CB}(t, t_f) - U_{CB}(t_i, t_f) = 2\pi s_o \left\{ \left[\ell(t) - \ell(t_i)\right] \int_{t}^{t_f} J(\tau) d\tau - \left[\ell(t_f) - \ell(t)\right] \int_{t_i}^{t} J(\tau) d\tau \right\}, \quad (25)$$

which is, in general, non-null.

In the particular case that the influx is constant, i.e. $J(\tau) = J$ in the interval $(t_i, t_f)$, the last expression (25) reduces to:

$$U_{CB}(t_i, t) + U_{CB}(t, t_f) - U_{CB}(t_i, t_f) =$$
$$= 2\pi s_o J \left\{ \left[\ell(t) - \ell(t_i)\right](t_f - t) - \left[\ell(t_f) - \ell(t)\right](t - t_i) \right\}, \qquad (26)$$
$$= 2\pi s_o J \left\{ \ell(t)(t_f - t_i) - \ell(t_i)(t_f - t) - \ell(t_f)(t - t_i) \right\}$$

which is also, in general, non-null. Moreover, if we define the function

$$g(t) = \ell(t)(t_f - t_i) - \ell(t_i)(t_f - t) - \ell(t_f)(t - t_i) \text{ in the interval } (t_i, t_f),$$

we can see immediately that the derivative of $g(t)$ is given by $\overset{\bullet}{g}(t) = \overset{\bullet}{\ell}(t)(t_f - t_i) + \ell(t_i) - \ell(t_f)$ and therefore $\overset{\bullet}{g}(t) = 0$ if and only if $\overset{\bullet}{\ell}(t) = \dfrac{\ell(t_f) - \ell(t_i)}{t_f - t_i} = \text{Const.}$ in the interval $(t_i, t_f)$, that is, the law of root growth must be linear, and then the function $g(t) = g_0 = \text{Const.}$ in the interval $(t_i, t_f)$ and therefore the expression (26) reduces to

$$U_{CB}(t_i, t) + U_{CB}(t, t_f) - U_{CB}(t_i, t_f) = 2\pi s_0 J g_0 = \text{Const.}$$

in the interval $(t_i, t_f)$. For this reason, for any law of root grow, different to the linear case, the expression (25) is not null.

In the particular case that the influx is constant, i.e. $J(\tau) = J$ in the interval $(t_i, t_f)$, and the law of root growth is linear, i.e., $\ell(t) = \ell(t_i) + k(t - t_i)$ in the interval $(t_i, t_f)$, then the last expression (26) reduces to:

$$
\begin{aligned}
U_{CB}(t_i, t) + U_{CB}(t, t_f) - U_{CB}(t_i, t_f) = \\
= 2\pi s_o J \left\{ \left[ \ell(t_i) + k(t - t_i) \right](t_f - t_i) - \ell(t_i)(t_f - t) - \left[ \ell(t_i) + k(t_f - t_i) \right](t - t_i) \right\} \\
= 2\pi s_o J \left\{ \ell(t_i)(t_f - t_i) + k(t - t_i)(t_f - t_i) - \ell(t_i)(t_f - t) - \ell(t_i)(t - t_i) - k(t_f - t_i)(t - t_i) \right\} = 0
\end{aligned}
\tag{27}
$$

Thus, we have proved analytically that, for the Claassen-Barber formula (18) the condition (22) is not satisfied, in general, except when the influx J is constant and the length of root grows linearly with the time.

Similarly, for the Cushman formula (19), we can simplify the original expression to:

$$
\begin{aligned}
U_C &= 2\pi s_o \left[ \ell(t_i) \int_{t_i}^{t_f} J(s)ds + \int_{t_i}^{t_f} \underbrace{\left[ \int_{t_i}^{t_i + t_f - \tau} J(s)ds + \right]}_{f(\tau)} \overset{\bullet}{\ell}(\tau)d\tau \right] \\
&= 2\pi s_o \left[ \ell(t_i) \int_{t_i}^{t_f} J(s)ds + f(\tau)\ell(\tau) \Big|_{\tau = t_i}^{\tau = t_f} - \int_{t_i}^{t_f} \underbrace{f'(\tau)}_{=-J(t_i + t_f - \tau)} \ell(\tau)d\tau \right] \\
&= 2\pi s_o \left[ \underbrace{\ell(t_i) \int_{t_i}^{t_f} J(s)ds}_{f(t_i)} + \underbrace{f(t_f)}_{=\int_{t_i}^{t_f} J(s)ds=0} \ell(t_f) - \underbrace{f(t_i)\ell(t_i)}_{=0} + \int_{t_i}^{t_f} J(t_i + t_f - \tau)\ell(\tau)d\tau \right] \\
&= 2\pi s_o \int_{t_i}^{t_f} J(t_i + t_f - \tau)\ell(\tau)d\tau
\end{aligned}
\tag{28}
$$

Now, in order to decide if $U_C$ verifies (22) we compute:

$$
\begin{aligned}
\frac{U_C(t_i, T) + U_C(T, t_f) - U_C(t_i, t_f)}{2\pi s_o} = \\
= \int_{t_i}^{t} J(t_i + t - \tau)\ell(\tau)d\tau + \int_{t}^{t_f} J(t + t_f - \tau)\ell(\tau)d\tau - \int_{t_i}^{t_f} J(t_i + t_f - \tau)\ell(\tau)d\tau \\
= \int_{t_i}^{t} J(t_i + t - \tau)\ell(\tau)d\tau + \int_{t}^{t_f} J(t + t_f - \tau)\ell(\tau)d\tau - \int_{t_i}^{t} J(t_i + t_f - \tau)\ell(\tau)d\tau - \int_{t}^{t_f} J(t_i + t_f - \tau)\ell(\tau)d\tau \\
= \int_{t_i}^{t} \left[ J(t_i + t - \tau) - J(t_i + t_f - \tau) \right]\ell(\tau)d\tau + \int_{t}^{t_f} \left[ J(t + t_f - \tau) - J(t_i + t_f - \tau) \right]\ell(\tau)d\tau
\end{aligned}
\tag{29}
$$

that is

$$
\begin{aligned}
U_C(t_i, T) + U_C(T, t_f) - U_C(t_i, t_f) = \\
= 2\pi s_o \left\{ \int_{t_i}^{t} \left[ J(t_i + t - \tau) - J(t_i + t_f - \tau) \right]\ell(\tau)d\tau + \int_{t}^{t_f} \left[ J(t + t_f - \tau) - J(t_i + t_f - \tau) \right]\ell(\tau)d\tau \right\}
\end{aligned}
\tag{30}
$$

which is, in general, non-null.

In the particular case that the influx is constant, i.e. $J(\tau) = J$ in the interval $(t_i, t_f)$, the last expression (30) vanishes because the two brackets in the previous expression are null. Therefore, when J is constant then $U_C$ verifies expression (22) regardless of the law of root growth.

Finally, the simplified formula Reginato-Tarzia (21) and its original version (20) obviously always verify condition (22) because of the linearity of the integral, that is:

$$\int_{t_i}^{t} J(t)\ell(t)d\tau + \int_{t}^{t_f} J(t)\ell(t)d\tau = \int_{t_i}^{t_f} J(t)\ell(t)d\tau \qquad (31)$$

Thus, the formula of Reginato-Tarzia always verifies the condition (22) whatever be the representative functions for the influx J(t) and the root length l(t). The obtained results show that the Claassen-Barber and Cushman cumulative nutrient uptake formulas do not satisfy, in general, the mass balance condition (22). In summary:

• The Claassen-Barber formula (18) only verifies the mass balance when the influx is constant (high concentrations) and the root grows linearly.

• The Cushman formula (19) verifies the mass balance when the influx is constant regardless of the law of growth.

• The Reginato-Tarzia formula (20) and his simplified version (21) are formulas which verify always the mass balance whatever be the representative functions for the influx and the law of root growth.

**APPENDIX V:** In order to compare observed and simulated influx on the root surface, a more adequate averaged influx is defined as (15). With this expression and taking into account the cumulative nutrient uptake formula (21) we obtain the experimental formula of Williams (1948). Thus, replacing (21) in (15) results:

$$\overline{J} = \frac{\int_{t_i}^{t_f} J(t)\ell(t)d\tau}{\int_{t_i}^{t_f} \ell(t)d\tau} = \frac{\Delta U / 2\pi s_o}{\int_{t_i}^{t_f} \ell(t)d\tau}.$$

For the exponential growth case $\left[\ell(t) = \ell_o e^{k(t-t_o)}\right]$ we have:

$$\int_{t_i}^{t_f} \ell(t)d\tau = \frac{(\ell_f - \ell_i)}{\ln\left(\ell_f / \ell_i\right)}(t_f - t_i).$$

Therefore, replacing this last expression in $\overline{J}$ we obtain the formula of Williams for the

exponential case:

$$2\pi s_o \overline{J} = \frac{\Delta U \ln\left(\ell_f/\ell_i\right)}{\left(\ell_f - \ell_i\right)\left(t_f - t_i\right)}$$

where the left hand side is given in mol/cm-s.

Similarly, for the linear growth case $\left[\ell(t) = \ell_o + k(t - t_o)\right]$ we have:

$$\int_{t_i}^{t_f} \ell(t) d\tau = \frac{\left(\ell_f + \ell_i\right)}{2}\left(t_f - t_i\right).$$

Therefore, replacing this last expression in $\overline{J}$ we obtains the formula of William for the linear case:

$$2\pi s_o \overline{J} = \frac{2\Delta U}{\left(\ell_f + \ell_i\right)\left(t_f - t_i\right)}$$

Thus, cumulative uptake formula (21), the redefinition of averaged influx (15) and the experimental formula of Williams are well posed. We remark that the formula of Williams is a consequence of our definition of weight averaged influx (15) for all root growth law expressions. This result cannot be obtained by using the temporal averaged influx except in the constant influx case (for high concentrations). This result has been partially corroborated by an experimental work of Silberbush and Gbur (1994).

Then, it is not convenient to choose the simulated temporal averaged influxes as an indicator of a good prediction for the cumulative nutrient uptake when the concentration are low, because by equation (21) the cumulative nutrient uptake U is proportional to the influx J only when this is constant. Some papers in the literature use erroneously the simulated temporal averaged influxes as a good prediction for low concentrations (Stritsis et al, 2014, Samal et al., 2010, and Singh et al. 2003).

## ACKNOLEDGEMENTS

This paper was partially sponsored by the project PIP No. 0534 of CONICET – Universidad Austral (Rosario, Argentina), and SECYT - UNRC No. 18/C427 (Río Cuarto, Argentina). The authors thank deeply Prof. Norbert Claassen for their valuable suggestions related to the successes and failures of the nutrient uptake models.

**REFERENCES**

Abbès, C. Parent, L. E. & Robert, J. L. (1996) Mechanistic Modeling of Coupled Ammonium and Nitrate Uptake by Onions Using the Finite Element Method, *Soil Sci. Soc. Am. J.* **60,** 1160-1167

Alexiades, V. & Solomon, A.D. (1993) Mathematical Modeling of Melting and Freezing Processes, Hemisphere, Washington, DC, Taylor & Francis, Washington, USA.

Barber, S.A. & Cushman, J. (1981) Nitrogen uptake model for agronomic crops in: I.R. Iskandar, Ed., Modeling wastewater renovation-land treatment, Wiley-Interscience, New York, 382-409

Brewster, J.L., Bhat K.K.S. & Nye P.H. (1976) The possibility of predicting solute uptake and plant growth response from independently measured soil and plant characteristics. V. The growth and phosphorus uptake of rape in soil at a range of phosphorus concentrations and a comparison of results with the predictions of a simulation model, *Plant Soil*, **44**, 295–328

Cannon, J.R. (1984) The One-Dimensional Heat Equation. Addison-Wesley, Menlo Park, CA.

Carslaw, H. & Jaeger, J.C. (1959) Conduction of Heat in Solids, 2nd Ed., Clarendon Press Oxford.

Claassen, N. & Barber, S.A. (1976) Simulation model for nutrient uptake from soil by a growing plant root system. *Agron. J.* **68**, 961–964.

Claassen, N., Syring, K.M. & Jungk, A. (1986) Verification of a mathematical-model by simulating potassium uptake from soil. *Plant and Soil* **95**, 209-220.

Comerford, N.B., Cropper, W.P.Jr., Hua, L., Smethurst, P.J., Van Rees K.C.J., Jokela, E.J. & Adams, F. (2006) Soil Supply and Nutrient Demand (SSAND): a general nutrient uptake model and an example of its application to forest management. *Can J. Soil Sci.* **86**, 665–673

Crank, J. (1984) Free and Moving Boundary Problems, Clarendon Press, Oxford.

Cushman, J.H. (1979) An analytical solution to solute transport near root surfaces for low initial concentrations: I. Equation development, *Soil Sc. Soc. Am. J.* **43**, 1087-1090.

Dunbabin, V.M., Postma, J.A., Schnepf, A., Pagès, L., Javaux, M., Wu, L., Leitner, D., Chen, Y.L., Rengel, Z. & Diggle, A.J. (2013) Modelling root–soil interactions using three–dimensional models of root growth, architecture and function, *Plant Soil*, **372**, 93–124

Heppell, J., Talboys, J.P., Payvandi, S., Zygalakis, K.C., Fliege, J., Withers, P.J.A., Jones, D.L. & Roose, T. (2014) How changing root system architecture can help tackle a reduction in soil phosphate (P) levels for better plant P acquisition, *Plant, Cell & Environment,* 1-33. (In Press)

Hinsinger, P. (2001) Bioavailability of soil inorganic P in the rhizosphere as affected by root induced chemical changes: a review. *Plant Soil*, **237**, 173-195.

Hinsinger, P.; Brauman, A.; Devau, N.; Gérard, F.; Jourdan, C.; Laclau, J.P.; Le Cadre. E.; Jaillard. B. & Plassard, C. (2011) Acquisition of phosphorus and other poorly mobile nutrients by roots. Where do plant nutrition models fail?, *Plant Soil,* **348**, 29-61

Huguenin-Elie, O., Kirk, G.J.D. & Frossard, E. (2003) Phosphorus uptake by rice from soil that is flooded, drained or flooded then drained. *European Journal of Soil Science*, **54,** 77-90.

Jonard, M., Augusto, L., Hanert, E., Achat, D.L., Bakker, M.R., Morel, C., Mollier, A. & Pellerin, S. (2010) Modeling forest floor contribution to phosphorus supply to maritime pine seedlings in two-layered forest soils *Ecological Modelling*, **221**, 927-935

Kelly, J.M., Barber, S.A. & Edwards, G.S. (1992) Modeling magnesium, phosphorus and potassium uptake by loblolly pine seedling using a Barber-Cushman approach *Plant and Soil* **139**, 209-218.

Li, H. & Comerford, N.B. (2000) SSAND Version 1.0, Release 1.06 (10-8-2001) *User's Guide*. University of Florida. Gainesville, FL.

Lin, W. & Kelly, J.M. (2010) Nutrient uptake estimates for woody species as described by the NST 3.0, SSAND, and PCATS mechanistic nutrient uptake models, *Plant Soil*, **335**, 199–212

Lu, S. & Miller, M.H. (1994) Prediction of phosphorus uptake by field-grown maize with the Barber-Cushman Model. *Soil Sci. Am. Soc. J.* **58**, 852–857

Lunardini, V.J. (1991) Heat Transfer with Freezing and Thawing. Elsevier, Amsterdam.

Mollier A., De Willigen P., Heinen M., Morel C., Scheinder A. & Pellerin S. (2008) A two dimensional simulation model of phosphorus uptake including crop growth and P response. *Eco. Mod.*, **210**, 453-464.

Nye, P.H. & Marriott, F.H.C. (1969) A Theoretical Study of the Distribution of Substances around Roots Resulting from Simultaneous Diffusion and Mass Flow, *Plant and Soil*, **30** (3), 459-472.

Oates, K. & Barber, S.A. (1987) NUTRIENT UPTAKE: A Microcomputer Program to Predict Nutrient Absorption from Soil by Roots. *Journal of Agronomic Education*, **16** (2), 65-68

Reginato, J.C., D.A. Tarzia & Cantero, A. (1990) On the free boundary problem for the Michaelis-Menten absorption model for root growth, *Soil Science*, **150**(4), 722-725

Reginato J.C., Palumbo M.C., Bernardo C.H., Moreno I. & Tarzia D.A. (2000) Modelling nutrient uptake using a moving boundary approach. Comparison with the Barber-Cushman model. *Soil Sci. Soc. Am. J.* **64**, 1363-1367.

Reginato J.C & Tarzia D.A. (2002) An alternative formula to compute the nutrient uptake for roots. *Commun. Soil Sci Plant Anal.* **33** (5&6), 821-830.

Roose, T., Fowler, A.C. & Darrah, P.R. (2001) A mathematical model of plant nutrient uptake. *J. Math. Biol.* **42**, 347–360

Roose T. & Kirk G.J.D. (2009) The solution of convection–diffusion equations for solute transport to plant roots. *Plant Soil*, **316**, 257–264

Schenk, M.K. & Barber, S.A. (1980) Potassium and phosphorus uptake by corn genotypes grown in the field as influenced by root characteristics, *Plant Soil*, **54**, 65–76

Schnepf, A., Schrefl, T. & Wenzel, W.W. (2002) The suitability of pde-solvers in rhizosphere modeling, exemplified by three mechanistic rhizosphere models. *J. Plant Nutr. Soil Sc.* **165**, -718

Samal, D., Kovar, J.L., Steingrobe, B., Sadana, U.S., Bhadoria, P.S. & Claassen, N. (2010) Potassium uptake efficiency and dynamics in the rhizosphere of maize (Zea mays L.), wheat (Triticum aestivum L.), and sugar beet (Beta vulgaris L.) evaluated with a mechanistic model, *Plant Soil*, **332**, 105–121

Silberbush, M. & Gbur, E.E. (1994) Using the Williams Equation to Evaluate Nutrient Uptake Rate by Intact Plants, *Agronomy J*. **86**, 107-110

Silberbush, M. (2013) Root Study: Why Is It behind Other Plant Studies?, American Journal of Plant Sciences, **4** (2), 198-203.

Singh, N., Bhadoria, P.B.S. & Rakshit, A. (2003) Simulation of Phosphorus Uptake by Peanut in Low Phosphorus Supplying Soil, *Ital. J. Agron.,* **7**, 1, 65-71

Smethurst, P.J., Mendham, D.S., Battaglia, M. & Misra, R. (2004) Simultaneous prediction of nitrogen and phosphorus dynamics in a Eucalyptus nitens plantation using linked CABALA and PCATS models. In: Borralho NMG (ed) Eucalyptus in a changing world. Proceedings of the IUFRO Conference, 11–15 October 2004. Aveiro, Portugal, 565–569

Steingrobe, B., Claassen, N. & Syring K.M. (2000) The effect of the function type for describing the soil buffer power on calculated ion transport to roots and nutrient uptake from the soil, *J. Plant Nutr. Soil Sci*. **163**, 459–465

Stritsis, C., Steingrobe, B. & Claassen; N. (2014) Cadmium Dynamics in the Rhizosphere and Cd Uptake of Different Plant Species Evaluated by A Mechanistic Model, *International Journal of Phytoremediation,* **16** (11), 1104-1118

Tarzia, D.A. (2000) A bibliography on moving-free boundary problems for the heat-diffusion equation. The Stefan and related problems, MAT - Serie A, Rosario, # 2 (with 5869 titles) 297 páginas. See: http://web.austral.edu.ar/cienciasEmpresariales-investigacion-mat-A-02.asp

Tarzia, D.A. (2011) Explicit and approximated solutions for heat and mass transfer problems with a moving interface, Chapter 20, In *Advanced Topics in Mass Transfer*, Mohamed El-Amin (Ed.), InTech Open Access Publisher, Rijeka, pp. 439-484. Available from: http://www.intechopen.com/articles/show/title/explicit-and-approximated-solutions-for-heat and-mass-transfer-problems-with-a-moving-interface

Tinker, P.B. & Nye, P.H. (2000) Solute Movement in the Rhizosphere, Oxford University Press, Oxford, England.

Jia-Xiang, X., Li-Gan, Z. & Wei-Min, Z. (1991) Mechanistic Model for Predicting $NO_3$-N Uptake by Plants and its Verification, *Pedosphere*, **1** (2), 97-108

Williams, R.F. (1946) The Physiology of Plant Growth with Special Reference to the Concept of Net Assimilation Rate, *Annals of Botany* **10** (37), 41-72

Williams, R.F. (1948) The Effects of Phosphorus Supply on the Rates of Intake of Phosphorus and Nitrogen and Upon Certain Aspects of Phosphorus Metabolism in Gramineous Plants, *Australian Journal of Scientific Research – Serie B – Biological Sciences* **1**, 333-361

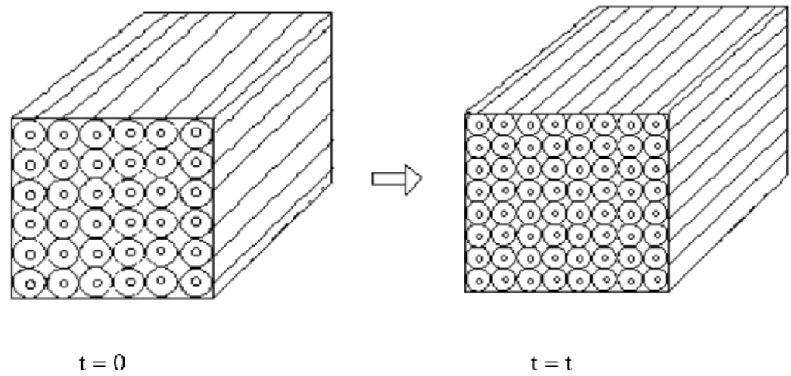

**Figure 1**. Homogeneous rooting in soil and its time evolution.

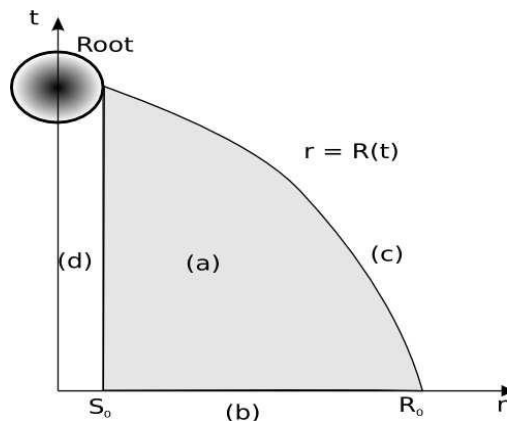

**Figure 2**. Scheme of the domain of validity of the proposed model in the plane r,t. The shaded zone (a) represents the domain where the equation (1) is valid. The line (b) is the spatial domain where initial value condition (2) is valid. The boundary condition (4) is valid on the curve c) (r = R(t)). In the line (d) the nutrient uptake condition (3) of the Michaelis-Menten type is used.

**Table 1.** Observed and predicted Cd influx by different crops to different soil Cd concentrations.

| Plant | Soil Conc. ($10^{-3}$ µmol cm$^{-3}$) | Influx of Cd ($10^{-16}$ mol cm$^{-2}$s$^{-1}$) | | | | | |
|---|---|---|---|---|---|---|---|
| | | Observed | Pred. NST 3.0 | Pred. / Obs. | Pred. MB-FE | Pred./ Obs. | Pred. Cum. Uptake MB-FE ( µmol) |
| Maize | 0.22 | 0.25 | 2.45 | 9.8 | *0.81* | *3.68* | *10.5* |
| Sunflower | 0.38 | 2.12 | 6.11 | 2.9 | *3.29* | *1.55* | *9.89* |
| Flax | 1.19 | 3.54 | 24.40 | 6.9 | *10.78* | *3.04* | *10.29* |
| Spinach | 0.48 | 7.55 | 12.00 | 1.6 | *9.78* | *1.29* | *4.72* |
| Maize+ | 0.74 | 1.64 | 7.73 | 4.7 | *1.9* | *1.16* | *22.73* |
| Sunflower+ | 1.80 | 5.56 | 25.90 | 4.7 | *10.2* | *1.83* | *32.42* |
| Flax+ | 4.59 | 10.98 | 82.20 | 7.5 | *44.87* | *4.08* | *33.18* |
| Spinach+ | 3.07 | 42.11 | 75.00 | 1.8 | *61.13* | *1.45* | *16.91* |

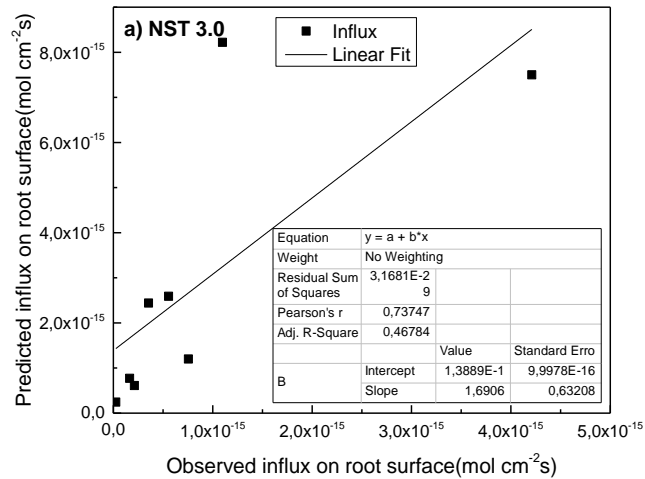

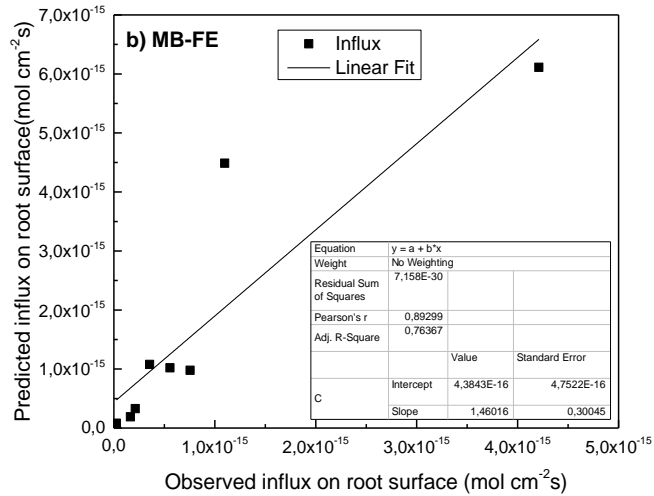

**Figure 3.** Observed and predicted influx by: a) Barber-Cushman model (NST 3.0 program), b) MB-FE model

**Table 2.** Observed and predicted K and P cumulative uptake by pine seedlings in soils with high nutrient concentrations using data extracted from literature [Kelly, 1992]

| Cumulative Uptake (µmol) | K | $U_p/U_o$ | P | $U_p/U_o$ |
|---|---|---|---|---|
| Observed | 6663 | | 1332 | |
| FB-NTS 3.0 | 6690 | 1.004 | 1320 | 0.99 |
| FB-NTS 3.0* | 6619.3 | 0.993 | 1304.9 | 0.979 |
| MB-FE | *6566.5* | *0.986* | *1273.3* | *0.957* |
| R.L.D. (180 days) | | $0.02 \rightarrow 0.20$ | | |

**Table 3.** Observed and predicted cumulative $NO_3$-N uptake by wheat in soils with high nutrient concentrations using data extracted from literature [Jia-Xiang et al., 1991]

| Cumulative Uptake (μmol) | Wheat | | | | | |
|---|---|---|---|---|---|---|
| | Red Soil | $U_p/U_o$ | Paddy soil | $U_p/U_o$ | Fluvo-aquic soil | $U_p/U_o$ |
| Observed | 189 | | 1263 | | 2205 | |
| FB-NTS 3.0 | 158 | 0.836 | 1470 | 1.164 | 2020 | 0.916 |
| FB-NTS 3.0* | 157.6 | 0.834 | 1466.7 | 1.161 | 2012.6 | 0.913 |
| MB-FE | *157.13* | *0.831* | *1461* | *1.15* | *2008.7* | *0.911* |
| R.L.D. | 0.62→0.7 (3 days) | | 0.83→1.3 (7 days) | | 0.67→1.5 (10 days) | |

**Table 4.** Observed and predicted cumulative P uptake by peanut without root hairs at different soil levels.

| Cumulative Uptake (μmol) | Peanut | | | | | | | | | |
|---|---|---|---|---|---|---|---|---|---|---|
| | 0 P | $U_p/U_o$ | +50 P | $U_p/U_o$ | +100 P | $U_p/U_o$ | +200 P | $U_p/U_o$ | +400 P | $U_p/U_o$ |
| Observed | 540 | | 640 | | 900 | | 1060 | | 1320 | |
| FB-NTS 3.0 | 1180 | 2.2 | 6230 | 9.7 | 18000 | 20 | 39900 | 37.6 | 44500 | 33.5 |
| FB-NTS 3.0* | 708.4 | 1.3 | 3704 | 5.8 | 11856 | 13 | 39922 | 37.6 | 43841 | 33.2 |
| MB-FE | *468.8* | *0.9* | *2211* | *3.4* | *5395* | *6* | *13338* | *12.6* | *42169* | *32* |
| R.L.D.(72 days) | 0.9→85 | | 0.9→100 | | 0.9→100 | | 1→107 | | 0.8→75 | |

**Table 5.** Observed and predicted P influx on root surfaceby peanut without root hairs at different soil P levels.

| Influx on root surface ($10^{-8}$ μmol cm$^{-1}$ s$^{-1}$) | Peanut | | | | | | | | | |
|---|---|---|---|---|---|---|---|---|---|---|
| | 0 P | $J_p/J_o$ | +50 P | $J_p/J_o$ | +100 P | $J_p/J_o$ | +200 P | $J_p/J_o$ | +400 P | $J_p/J_o$ |
| Observed | 3.44 | | 3.59 | | 5.15 | | 6.13 | | 10.11 | |
| FB-NTS 3.0 | 0.29 | 0.08 | 1.33 | 0.37 | 3.65 | 0.7 | 7.5 | 1.2 | 10.9 | 1.08 |
| MB-FE | *1.15* | *0.33* | *4.53* | *1.26* | *10.83* | *2.1* | *25.75* | *4.2* | *105.3* | *10.4* |

**Table 6.** Average predicted/observed ratio for influx on root surface and cumulative uptake obtained by NST 3.0, NST 3.0* and MB-FE

| $\overline{(J_p/J_o)}$ | $\overline{(J_p/J_o)}$ | $\overline{(U_p/U_o)}$ | $\overline{(U_p/U_o)}$ | $\overline{(U_p/U_o)}$ |
|---|---|---|---|---|
| NST 3.0 | MB-FE | NST 3.0 | NST 3.0* | MB-FE |
| 4.74 | *3.24* | 18.1 | 18.1 | *10.9* |

**Table 7.** Observed and predicted cumulative K uptake for maize, wheat, and sugar beet grown on a low K soil with (+K) and without (-K) fertilization

| Cumulative Uptake (μmol) | Maize | | | | Wheat | | | | Sugar beet | | | |
|---|---|---|---|---|---|---|---|---|---|---|---|---|
| | -K | $U_p/U_o$ | +K | $U_p/U_o$ | -K | $U_p/U_o$ | +K | $U_p/U_o$ | -K | $U_p/U_o$ | +K | $U_p/U_o$ |
| Observed | 678 | | 1633 | | 524 | | 759 | | 434 | | 1035 | |
| FB-NST 3.0 | 3180 | 4.7 | 24100 | 14.7 | 5410 | 10.3 | 7600 | 10.0 | 4490 | 10.3 | 44900 | 43.4 |
| FB-NST 3.0* | 1421 | 2.1 | 18445 | 11.3 | 2657 | 5.1 | 7412 | 9.8 | 3173 | 7.3 | 30968 | 30 |
| *MB-FE* | *380* | *0.56* | *1563* | *0.95* | *443* | *0.84* | *1769* | *2.3* | *4642* | *1.1* | *1809* | *1.7* |
| R.L.D | 0.9→22 (21 days) | | | | 0.4→27 (26 days) | | | | 0.1→18 (26 days) | | | |

**Table 8.** Observed and predicted root K influx on root surface for maize, wheat, and sugar beet grown on a low K soil with (+K) and without (-K) fertilization

| Influx on root surface ($10^{-7}$ μmol $cm^{-1}s^{-1}$) | Maize | | | | Wheat | | | | Sugar beet | | | |
|---|---|---|---|---|---|---|---|---|---|---|---|---|
| | -K | $J_p/J_o$ | +K | $J_p/J_o$ | -K | $J_p/J_o$ | +K | $J_p/J_o$ | -K | $J_p/J_o$ | +K | $J_p/J_o$ |
| Observed | 1.99 | | 3.87 | | 2.39 | | 3.22 | | 8.45 | | 19.0 | |
| FB-NST 3.0 | 1.27 | 0.64 | 4.33 | 1.12 | 1.77 | 0.68 | 3.9 | 1.21 | 2.64 | 0.31 | 15.1 | 0.8 |
| *MB-FE* | *1.67* | *0.84* | *2.48* | *0.64* | *1.52* | *0.58* | *8.28* | *2.6* | *2.7* | *0.31* | *7.1* | *0.38* |

**Table 9.** Average predicted/observed ratio of influx on root surface and cumulative uptake obtained by NST 3.0, NST 3.0* and MB-FE.

| $\overline{\left(J_p/J_o\right)}$ | $\overline{\left(J_p/J_o\right)}$ | $\overline{\left(U_p/U_o\right)}$ | $\overline{\left(U_p/U_o\right)}$ | $\overline{\left(U_p/U_o\right)}$ |
|---|---|---|---|---|
| NST 3.0 | MB-FE | NST 3.0 | NST 3.0* | MB-FE |
| 0.8 | *0.88* | 15.6 | 10.9 | *1.25* |

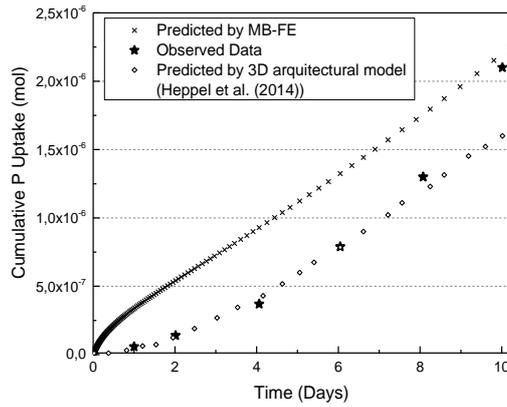

**Figure** 4. Observed, MB-FE and Heppel predicted values for the cumulative uptake of P by wheat seedlings over a 10 days period when grown in and low-P soils.

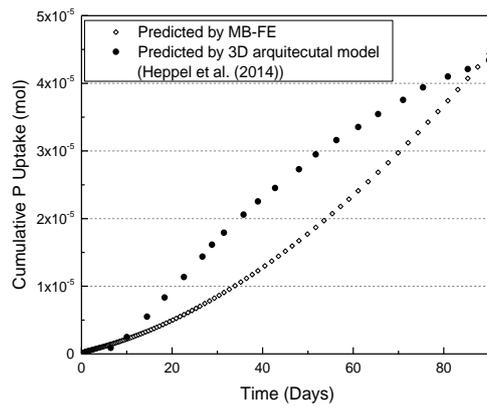

**Figure 5**. Predicted cumulative plant P acquisition by the MB-FE and the Heppel et al. model with an exponential branching distribution over a 90 days period when grown in and low-P soils.

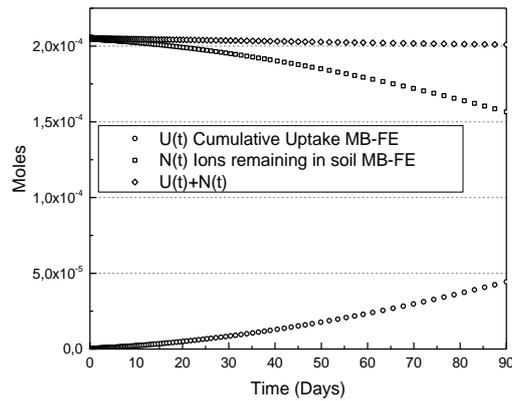

**Figure 6.** Mass balance for the cumulative P uptake and the P ions remaining with data of Heppel in soil the moving boundary model.

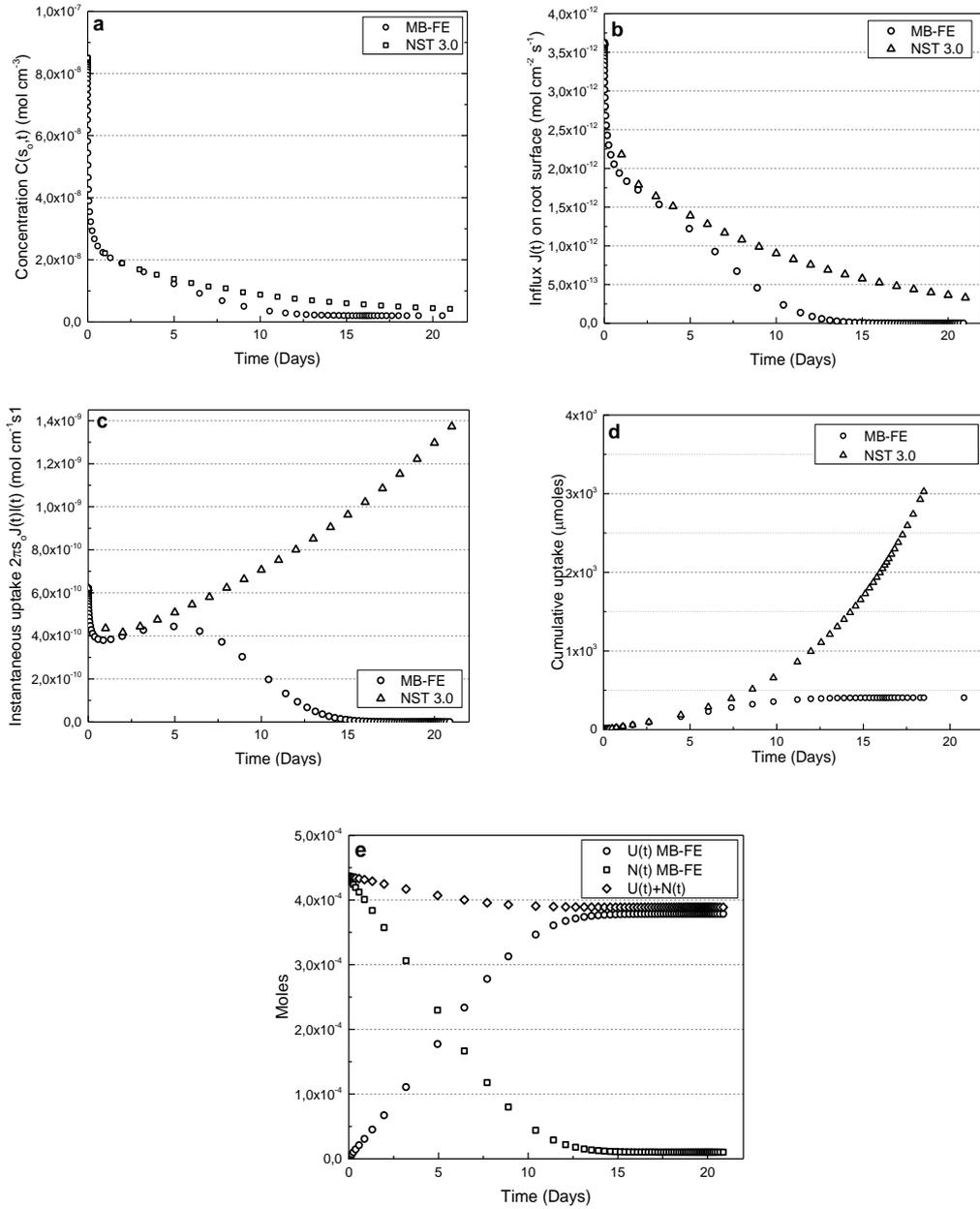

**Figure 7.** Comparison of obtained results between NST 3.0 and MB-FE models with data of Samal (2010) for maize without K addition. Concentration on root surface $C(s_o,t)$, b) influx on root surface $J[C(s_o,t)]$, c) instantaneous uptake $2\pi\,J(t)\,\ell(t)$, d) cumulative nutrient uptake $U(t)$ as a function of time, and e) corresponding mass balance for $U(t)$ and $N(t)$.